\documentclass[a4paper,12pt]{article}
\pdfoutput=1
\usepackage{feynmp-auto,expdlist}
\usepackage{amsmath, amsfonts, amssymb,mathbbol}
\usepackage{graphicx}
\usepackage{enumerate,cancel}
\usepackage{hyperref}
\usepackage{latexsym}
\usepackage{dsfont}
\usepackage{hepnicenames}
\usepackage{enumerate}
\usepackage{soul}
\usepackage[normalem]{ulem}
\usepackage{comment}
\usepackage{feynmp-auto}
\newcommand{\marrow}[5]{%
    \fmfcmd{style_def marrow#1
    expr p = drawarrow subpath (1/3, 3/4) of p shifted 12 #2 withpen pencircle scaled 0.5;
    label.#3(btex #4 etex, point 0.5 of p shifted 18 #2);
    enddef;}
    \fmf{marrow#1,tension=0}{#5}}

\usepackage{units}

\usepackage{soul}

\usepackage{mathrsfs,graphicx,rotating,amsmath,amsfonts,mathtools,booktabs,amssymb,wasysym}
\usepackage{hyperref}\usepackage{slashed}
\usepackage[nosort]{cite}
\usepackage[table,xcdraw,dvipsnames]{xcolor}
\usepackage{bm}
\usepackage{graphicx}
\usepackage{multirow,multicol}
\hypersetup{colorlinks,bookmarksopen,bookmarksnumbered,
	linkcolor=blus,pdfstartview=FitH,urlcolor=rossos,citecolor=verde}
\allowdisplaybreaks

\renewcommand{\[}{\left[}

\renewcommand{\vec}[1]{\boldsymbol{#1}}

\newcommand{\mio}[1]{}

\newcommand{\med}[1]{\langle #1\rangle}

\newcommand{\bpm}{\begin{pmatrix}}
\newcommand{\epm}{\end{pmatrix}}

\usepackage{mathrsfs}

\newcommand{\fig}[1]{~\ref{fig:#1}}

\newcommand{\One}{1\!\!\hbox{I}}
\renewcommand{\One}{\mathbb{1}}

\allowdisplaybreaks
\usepackage{multicol}
\usepackage{color}
\definecolor{rosso}{cmyk}{0,1,1,0.4}
\definecolor{rossos}{cmyk}{0,1,1,0.55}
\definecolor{rossoc}{cmyk}{0,1,1,0.2}
\definecolor{blu}{cmyk}{1,1,0,0.3}
\definecolor{blus}{cmyk}{1,1,0,0.6}
\definecolor{bluc}{cmyk}{1,1,0,0.1}
\definecolor{verde}{cmyk}{0.92,0,0.59,0.25}
\definecolor{verdec}{cmyk}{0.92,0,0.59,0.15}
\definecolor{verdes}{cmyk}{0.92,0,0.59,0.4}

\newcommand{\eq}[1]{~{\rm (\ref{eq:#1})}}

\newcommand{\GeV}{\,{\rm GeV}}
\newcommand{\TeV}{\,{\rm TeV}}

\newcommand{\Tr}{\,{\rm Tr}}
\newcommand{\diag}{\,{\rm diag}}

\newcommand{\beq}{\begin{equation}}
\newcommand{\eeq}{\end{equation}}
\newcommand{\mb}[1]{\mbox{\boldmath $#1$}}
\newcommand{\bea}{\begin{eqnarray}}
\newcommand{\eea}{\end{eqnarray}}
\newcommand{\be}{\begin{equation}}
\newcommand{\ee}{\end{equation}}
\font\tenrsfs=rsfs10 at 12pt
\font\sevenrsfs=rsfs7
\font\fiversfs=rsfs5
\newfam\rsfsfam
\textfont\rsfsfam=\tenrsfs
\scriptfont\rsfsfam=\sevenrsfs
\scriptscriptfont\rsfsfam=\fiversfs

\newcommand{\kb}[2]{|#1\rangle \langle #2|}
\newcommand{\kk}[1]{|#1\rangle}

\def\be#1\ee{\begin{equation}#1\end{equation}}
\def\bl#1\el{\begin{align}#1\end{align}}
\def\ba#1\ea{\begin{align*}#1\end{align*}}

\renewenvironment{thebibliography}[1]
{\begin{multicols}{2}[\section*{\refname}]%
		\@mkboth{\MakeUppercase\refname}{\MakeUppercase\refname}%
		\list{\@biblabel{\@arabic\c@enumiv}}%
		{\settowidth\labelwidth{\@biblabel{#1}}%
			\leftmargin\labelwidth
			\advance\leftmargin\labelsep
			\@openbib@code
			\usecounter{enumiv}%
			\let\p@enumiv\@empty
			\renewcommand\theenumiv{\@arabic\c@enumiv}}%
		\sloppy
		\clubpenalty4000
		\@clubpenalty \clubpenalty
		\widowpenalty4000%
		\sfcode`\.\@m}
	{\def\@noitemerr
		{\@latex@warning{Empty `thebibliography' environment}}%
		\endlist\end{multicols}}

\newcommand{\eV}{\,{\rm eV}}
\newcommand{\SU}{\,{\rm SU}}
\newcommand{\SO}{\,{\rm SO}}
\newcommand{\U}{\,{\rm U}}

\makeatletter
\font\ital=cmu10

\def\hhref#1{\href{http://arxiv.org/abs/#1}{arXiv:#1}}
\usepackage{xstring}
\newcommand{\hhrefq}[1]{\IfSubStr{#1}{:}{\href{http://inspirehep.net/search?ln=en&ln=en&p=#1&of=hb&action_search=Search&sf=&so=d&rm=&rg=25&sc=0}{InSpire:#1}}{\hhref{#1}}}

\def\art{\@ifnextchar[{\eart}{\oart}}
\def\eart[#1]#2#3#4#5#6{{\rm #2}, {\em #3 \bf #4} {\rm (#6) #5} ({\em #1})}
\def\article{\@ifnextchar[{\earticle}{\oarticle}}
\def\oarticle#1#2#3#4#5#6{{\rm #1}, {\ital `#6'}, {\rm #2 #3 (#5) #4}}
\def\earticle[#1]#2#3#4#5#6#7{{\rm #2}, {\ital `#7'}, {\rm #3 #4 (#6) #5}  [\hhrefq{#1}]}
\def\hepart[#1]#2{{\rm #2, \sl#1}}
\def\heparticle[#1]#2#3{#2, {\ital `#3'} [\hhrefq{#1}]}
\newcommand{\doi}[1]{\href{http://dx.doi.org/#1}{[link]}}

\newcommand{\hhrefqq}[1]{\IfBeginWith{#1}{10.}{\href{https://doi.org/#1}{doi:#1}}{\hhrefq{#1}}}

\renewenvironment{thebibliography}[1]
{\begin{multicols}{2}[\section*{\refname}]%
		\@mkboth{\MakeUppercase\refname}{\MakeUppercase\refname}%
		\list{\@biblabel{\@arabic\c@enumiv}}%
		{\settowidth\labelwidth{\@biblabel{#1}}%
			\leftmargin\labelwidth
			\advance\leftmargin\labelsep
			\@openbib@code
			\usecounter{enumiv}%
			\let\p@enumiv\@empty
			\renewcommand\theenumiv{\@arabic\c@enumiv}}%
		\sloppy
		\clubpenalty4000
		\@clubpenalty \clubpenalty
		\widowpenalty4000%
		\sfcode`\.\@m}
	{\renewcommand{\@noitemerr}
		{\@latex@warning{Empty `thebibliography' environment}}%
		\endlist\end{multicols}}

\newcounter{alphaequation}[equation]
\renewcommand{\thealphaequation}{\theequation\hbox to
	0.6em{\hfil\alph{alphaequation}\hfil}}
\definecolor{Gray}{gray}{0.95}

\usepackage{tocloft}
\begin{document}
\thispagestyle{empty}
\begin{center}  
{\LARGE\bf\color{rossos} New physics in spin entanglement} \\
\vspace{0.6cm}
 {\bf Mateusz Duch}, {\bf Alessandro Strumia} and {\bf Arsenii Titov}  \\[6mm]
{\it Dipartimento di Fisica, Universit\`a di Pisa, Italia}\\[1mm]

\vspace{0.5cm}
{\large\bf Abstract}
\begin{quote}\large
We propose a theory that preserves spin-summed 
scattering and decay rates at tree level while affecting particle spins.
This is achieved by breaking the Lorentz group in a non-local way that tries avoiding  stringent constraints,
for example leaving unbroken the maximal sub-group SIM(2).
As a phenomenological application, 
this new physics can alter the spins of top-antitop pairs 
(and consequently their entanglement)
produced in $pp$ collisions without impacting their rates.
Some observables affected by loops involving top quarks with modified entanglement
receive corrections.
\end{quote}
\end{center}
\setcounter{page}{1}
\tableofcontents
\newpage

\section{Introduction}
The ATLAS~\cite{ATLAS} and CMS~\cite{2406.03976} experiments at the Large Hadron Collider confirm that proton-proton ($pp$)  collisions at energy $\sqrt{s}\approx 13\TeV$ produce 
$ t \bar t$ pairs of top quarks with an entangled spin structure as predicted by the
Standard Model. 
The Standard Model predicts that the~dominant~top~pair production process at the LHC is from  gluons, $gg\to t \bar t$.
Since gluons have spin~1, two gluons produce
maximally entangled top pairs in the spin-singlet state,
$\kk{t_\uparrow \bar{t}_\downarrow}-\kk{t_\downarrow \bar{t}_\uparrow}$,
around the kinematical threshold where tops are non-relativistic~\cite{hep-ph/9512292,2402.07972}.
Measuring top quark spins necessitates a sophisticated experimental analysis, 
as top quark spin is indirectly inferred from the angular distributions of the leptons resulting from weak top quark decays $t \to b \bar\ell \nu$,
see~fig.\fig{FeynEntanglementLHC}.

\smallskip

From a theoretical perspective, this experimental result comes as no surprise.
Indeed entanglement is a well known phenomenon~\cite{Einstein:1935rr,Schr35} at the 
basis of quantum mechanics that is currently receiving renewed attention.
 On one hand, the emergence of novel physics that alters entanglement seems implausible
 as any modification to quantum mechanics risks 
spoiling the quantum consistency of the Standard Model and, consequently, various observed quantities influenced by top-quark loops. 
On the other hand, conventional new physics scenarios 
(e.g., in the form of effective operators such as a chromo-magnetic dipole 
of the top~\cite{2211.10513,2203.05619,2208.11723,2210.09330,2402.07972})
could easily impact entanglement by modifying the $gg \to t \bar t$ and/or $t\to b \bar\ell \nu$ processes. 
However, such modifications would also influence the rates summed over spins,
 which have been tested  in multiple  ways  more extensively and more directly than entanglement itself.
For example, modifying QCD would also affect the $gg \to t\bar t$ total cross section,
and its angular and energy distributions summed over spins.
So the following issues arise:
\begin{enumerate}
\item Do collider measurements of entanglement at high energy test any new physics theory?
\listpart{In section~\ref{th}
we answer positively, proposing an unusual kind of new physics that
does not affect differential rates summed over spins by a single bit,
while affecting the qubit: spins and their entanglement get modified.
In section~\ref{tree} we show that this new physics
can modify the top spin entanglement observable measured by ATLAS and CMS.
This allows us to move beyond tree-level observables and address the next big issue:
}
\item Is new physics that affects top quark spin entanglement 
allowed by constraints on processes mediated by virtual top quarks at the loop level,
as observed in Higgs physics, electro-weak physics, flavour physics, low-energy physics?
\end{enumerate}
This issue is addressed in section~\ref{loop}
and is particularly important because the new theoretical ingredient we introduced involves a non-local breaking of Lorentz invariance.
Different unbroken Lorentz sub-groups are explored, and the maximal SIM(2) sub-group alleviates most unwanted loop effects.%
\footnote{Theories invariant under SIM(2) were previously discussed
to address neutrino mass~\cite{hep-ph/0601236,hep-ph/0605036} and dark~matter~\cite{1008.0436}.}
Conclusions and a summary of results are provided in section~\ref{concl}.


\section{A theory for new physics in spin entanglement}\label{th}
We propose a modification of the usual relativistic Quantum Field Theory
that, at tree level, leaves unaffected differential cross sections summed over spins, 
while modifying spins and thereby their entanglement.
Our theory can be defined as follows:
we replace a Dirac spinor field $\Psi(x)$  (for example the field describing top quarks) with
\beq \label{eq:PsiTilde}
\tilde\Psi(x) =W(i\partial) \Psi(x)
=\int \frac{d^3p}{(2\pi)^3 }\frac{1}{\sqrt{2E_p}}
\sum_{s} 
\left( \tilde{u}_{s}(p) a_{s \boldsymbol{p}}  e^{-i p\cdot x}
+  \tilde{v}_{s}(p) b_{s\boldsymbol{p}}^\dagger \, e^{i p\cdot x}
\right) \eeq
that differs from the standard expression because spinors are
replaced by  tilted ones 
\beq \label{eq:tiltedspinors} \tilde{u}_s(p) \equiv W(p) u_{s}(p)  ,\qquad
\tilde{v}_s(p) \equiv W(-p) v_{s}(p) \eeq
where $W(p)$ is a $4\times 4$ matrix in spinor space
that introduces a relative orientation between spin in space-time  and in spinor space,
and $E_p^2 =\mb{p^2}+M^2$.
Different choices of $W(p)$ give different theories.
If performed everywhere in the action, the 
\beq \label{eq:Psitildep} \Psi \to \tilde{\Psi} = W \Psi\eeq
replacement is a field redefinition that leaves physics invariant.
The transformation introduces new physics
if performed differently in different terms (for example, if performed in a sub-set of terms).
\begin{itemize}
\item We perform the $\Psi\to \tilde{\Psi}$ transformation of eq.\eq{Psitildep} in the Dirac kinetic term of $\Psi$,
that, in general, becomes a new operator $\tilde{\cal K}\neq {\cal K} \equiv \slashed{p}-M$:
\beq 
 {\bar{\Psi}} {\cal K} \Psi \to
  {\bar{\Psi}}\overline{W} (\slashed{p}-M)W \Psi \equiv
{\bar{\Psi}}  \tilde{\cal K} \Psi \eeq
where $\overline{W} = \gamma_0 W^\dagger \gamma_0$.
The fermion propagator $\Pi(p) = i /(\slashed{p}-M)$ gets modified into
\beq \label{eq:Pitilde}
\tilde{\Pi}(p) ={W}^{-1}(p) \cdot \Pi(p) \cdot \overline{W}^{-1}(p).\eeq
We will assume $W(p)$ has a non-local form such that
\beq \label{eq:W1}
\overline{W} W = \One,\qquad \overline{W} \slashed{p}W = \slashed{p}\eeq
because, as discussed later, this guarantees that spin-summed tree-level cross sections remain invariant.
Then $\tilde{\cal K}$ and its inverse, the propagator $\tilde{\Pi}$, keep the usual Dirac form.

\item Vertices that involve $\Psi$ get modified by
extra $W$ matrices. For example,
if $\Psi$ is the top quark field,
we can assume that $\tilde\Psi$  transforms in the standard way under the local QCD gauge symmetry.
Then the  top/gluon interaction
$g_3 g^a_\mu(q_g) T^a_{ij} \bar{\tilde{t}}_i(q_1) \gamma^\mu \tilde{t}_j(q_2) $
with $q_1+q_2 + q_g = 0$
gives the modified $g\bar t  t $ vertex:\\
\vspace{0.2cm}
\setlength{\unitlength}{1mm}
 \begin{fmffile}{gtt}
\beq
\begin{gathered}
    \begin{fmfgraph*}(22,14)
\fmfleft{i}
\fmfright{b,a}
\fmflabel{$g_\mu,a$}{i}
\fmflabel{$t,i$}{a}
\fmflabel{$\bar{t},j$}{b}
\fmf{gluon}{i,v1}
\fmf{plain_arrow}{v1,a}
\fmf{plain_arrow}{b,v1}
\fmfdot{v1}
\marrow{d}{up}{top}{$q_1$}{v1,a}
\marrow{e}{down}{bot}{$q_2$}{v1,b}
    \end{fmfgraph*}
\end{gathered}
\quad
=    ~~
-i g_3   T^a_{ij}  \, \overline{W} (q_1)\gamma^\mu W(- q_2).
\label{eq:QCDvertex}
\eeq
\end{fmffile}

\end{itemize}
Amplitudes will remain unmodified because, in Feynman diagrams, the inverse $W(p)$ factors arising from propagators in eq.\eq{Pitilde}
cancel with the $W(p)$ factors arising from vertices.
Next, the extra cancellation of $W(p)$ factors within propagators will ensure that spin-summed cross sections remain unmodified.
On the other hand, spins get rotated by $W(p)$ provided that rotational invariance and thereby Lorentz invariance is broken by $W(p)$.
We recall that a Lorentz transformation $ S\stackrel{\Lambda}{\to} S'$ from a reference frame $S$ to a frame $S'$
acts on coordinates as $x'=\Lambda x$ and in Hilbert space as the operator ${\cal U}_\Lambda$ 
\beq 
 {\cal U}_\Lambda a_{s\boldsymbol{p}}  {\cal U}^{-1}_\Lambda = 
 \sqrt{\frac{E_{p'}}{E_p}}\, D_{ss'}^{-1} ( \Lambda,p)\, a_{s'\boldsymbol{p}' },\qquad
 p' = \Lambda p
 \eeq
where $D$ is the unitary spin-representation of the Wigner spatial rotation associated to $\Lambda$ and $p$.
So the field $\tilde\Psi$ Lorentz transforms into
\beq \label{eq:LorentzPsi}
{\cal U}_\Lambda \tilde\Psi(x) {\cal U}^{-1}_\Lambda=\Lambda_{1/2}^{-1} 
\int \frac{d^3 p'}{(2\pi)^3} \frac{1}{\sqrt{2 E_{p'}}} \sum_{s'}
 \left[W'(p') a_{s'\boldsymbol{p}'}  u_{s'}(p')  e^{-ip'\cdot  x'} 
 + W'(-p') b_{s'\boldsymbol{p}'}^{\dagger} v_{s'}(p')  e^{ip'\cdot x'}\right] 
\eeq
showing that, in addition to the usual $\Lambda_{1/2}$ Lorentz matrix acting on spinors,
the extra factor $W(p)$ transforms into
\beq \label{eq:LorentzW}
W(p)\stackrel{\Lambda}\to W'(p') =  \Lambda_{1/2} W(p=\Lambda^{-1} p') \Lambda_{1/2}^{-1}\,.\eeq
Different forms of $W(p)$ break the Lorentz group down to different sub-groups $H$.
Furthermore, the space-time discrete symmetries parity P, charge conjugation C, and time reversal T
transform $W(p)$ as
\beq
\begin{array}{rcl}
 W(p) \xrightarrow{\rm P} & W'({\rm P}\,p) &=  \gamma^0 W(p) \gamma^0\,,   \cr
 W(p) \xrightarrow{\rm C}& W'(-p)& = C {W}^\ast(p) C  \cr
 W(p) \xrightarrow{\rm T}& \!\!W'({\rm P}\,p) \!\!& =  \gamma^3\gamma^1 W^\ast(p) \gamma^1\gamma^3 
\end{array}\label{eq:CPT}
\eeq
where ${\rm P}\,(p_0,\mb{p}) = (p_0, -\mb{p})$ and $C=-i\gamma^2$.
In the next 3 sections we will consider three possible symmetry breaking patterns:
\begin{enumerate}[a)]
\item  In section~\ref{SO21} we naively break Lorentz to SO(2,1).

\item  In section~\ref{rot} we try breaking rotational invariance while
keeping the associated breaking of boosts as mild as possible.

\item In section~\ref{SIM} we break Lorentz down to its phenomenologically acceptable maximal sub-group $H={\rm SIM}(2)$ 
that contains all boosts (combined with rotations), such that a special orientation is introduced without that any
rest frame becomes special.
\end{enumerate}

\subsection{Lorentz broken to SO(2,1)}\label{SO21}
The Lorentz group contains 3 rotation generators $J_{x,y,z}$ and 3 boost generators $K_{x,y,z}$.
As  a first naive attempt we break Lorentz to the sub-group $H$ that leaves invariant the space-like unit vector
\beq n^\mu = (0,\mb{n}) = (0,0,0,1).\eeq
Without loss of generality we can orient the $z$ axis along $\mb{n}$,
so that $H=\SO(2,1)$ is spanned by the three $J_z, K_x, K_y$ generators.
We assume the following two forms of $W(p)$:
\begin{enumerate}
\item Two equal rotations with angle $\delta/2$ around the $\mb{n}$ axis:
\beq \label{eq:WSO21a}
W(p)=\exp\left(i \delta  \frac{[ \slashed{n},\slashed{p}]\gamma_5}{8M_p}\right),
\qquad W(0)=\diag (e^{i\delta/4},e^{-i\delta/4},e^{i\delta/4},e^{-i\delta/4}). \eeq
We define $M_p^2 \equiv p \cdot p$ as $M_p \ge 0$ if $p^0>0$
and as $M_p \le0$ if $p^0<0$, such that $W(p)=W(-p)$ and a common $W$, denoted as $W(0)$, 
rotates in the same way spins of particles and anti-particles produced at rest.
This $W(p)$ is  invariant under C, P, and T, see eq.~(\ref{eq:CPT}) and table~\ref{tab:CPT}.
The choice $M_p>0$ for all $p^0$ would give opposite rotations, more simply obtained in the second example below.


\item Two opposite rotations with angle $\delta/2$ around the $\mb{n}$ axis:
\beq  \label{eq:WSO21b}
W(p)=\exp(-i \delta \,\slashed{n} \gamma_5/4),\qquad W(0)=\diag (e^{i\delta/4},e^{-i\delta/4},e^{-i\delta/4},e^{i\delta/4}). \eeq
This $p$-independent $W$
respects P and T but breaks C. 
Particles and anti-particles produced with the same spin and momenta acquire opposite phases therefore the parameter $\delta$ is odd under C, CP, and CPT (see table~\ref{tab:CPT}).

\end{enumerate}
Given their exponential form, both $W(p)$
satisfy eq.\eq{W1} and consequently leave unaffected the spin-summed tree-level cross sections.
Both $W(p)$ rotate spins and thereby affect spin entanglement.
However, at the loop level, these theories are subject to strong bounds on Lorentz breaking.
Therefore in the next two sections, we consider less problematic forms of Lorentz symmetry breaking.

\begin{table}[t]
 \centering
 \begin{tabular}{|cl|ccc|ccc|c|}
 \hline
 Lorentz broken to & Theory with $W(p)$ given by & P & T & C & PT & CP & CT & CPT \\
\hline\hline \rowcolor[rgb]{0.95,0.97,0.97}
 SO(2,1) & eq.\eq{WSO21a}, same rotation & $+$ & $+$ & $+$ & $+$ & $+$ & $+$ & $+$ \\
\rowcolor[rgb]{0.95,0.97,0.97}
SO(2,1)     &eq.\eq{WSO21b}, opposite rotation  & $+$ & $+$ & $-$ & $+$ & $-$ & $-$ & $-$ \\
 \rowcolor[rgb]{0.97,0.97,0.93}
  \hline \rowcolor[rgb]{0.99,0.97,0.97}
 SO(2)     & eq.\eq{W0rot}, same rotation & $+$ & $+$ & $+$ & $+$ & $+$ & $+$ & $+$ \\
 \rowcolor[rgb]{0.99,0.97,0.97}
SO(2)     &eq.\eq{Wn}, opposite rotation  & $+$ & $+$ & $-$ & $+$ & $-$ & $-$ & $-$ \\
 \rowcolor[rgb]{0.97,0.97,0.93}\hline
SIM(2)    &eq.\eq{SIMW0}, boost & $\color{rossos}\times$ & $\color{rossos}\times$ & $+$ & $+$ & $\color{rossos}\times$ & $\color{rossos}\times$ & $+$ \\
\rowcolor[rgb]{0.97,0.97,0.93}
SIM(2)    &eq.\eq{SIMW1}, same rotation  & $\color{rossos}\times$ & $\color{rossos}\times$ & $+$ & $+$ & $\color{rossos}\times$ & $\color{rossos}\times$ & $+$ \\
 \rowcolor[rgb]{0.97,0.97,0.93}
SIM(2)    &eq.\eq{SIMW2}, opposite rotation  & $\color{rossos}\times$ & $\color{rossos}\times$ & $-$ & $+$ & $\color{rossos}\times$ & $\color{rossos}\times$ & $-$ \\
 \hline
 \end{tabular}
  \caption{\em Transformation properties of the various $W(p,\delta) $ under space-time discrete symmetries {\rm S}.
A {$\color{rossos}\times$} means that the symmetry is broken.
A {$+$} means that $W(p,\delta) \xrightarrow{\rm S} W({\rm S}\,p,\delta)$ is invariant.
A $-$ means that $W(p,\delta) \xrightarrow{\rm S} W({\rm S}\,p,-\delta)=W({\rm S}\,p,\delta)^{-1}$, namely that
$\delta$ is odd under {\rm S}.
}
 \label{tab:CPT}
\end{table}

\subsection{Lorentz broken to SO(2)}\label{rot}
We modify the theory of section~\ref{SO21} by trying to make the
unavoidable breaking of boosts as mild as possible.
We start by considering a $\Psi$ particle at rest (denoted as 0) in the reference frame $S$.
We assume that, in the Dirac basis, 
\beq W(0)=\diag (U, \epsilon V^* \epsilon^{-1})\eeq
where  $\epsilon$ is the $2\times 2$ anti-symmetric tensor and $U,V$ are two $2\times 2$ matrices.
The spinor polarizations can be written in terms of their values at rest
as $u(p) = \Lambda_p u(0)$ and $v(p) = \Lambda_p v(0)$,
where \beq \label{eq:Lambdap}
\Lambda_p = 
 \frac{M_p  + \slashed{p}\gamma^0}{\sqrt{2M_p(p_0 + M_p)}}.\eeq
is the Lorentz transformation that brings to rest a time-like 
vector  $p_\mu$,
$\Lambda^{-1}_p \slashed{p}\Lambda _p= p \gamma_0$, 
with $M_p$ defined below eq.\eq{WSO21a}.
Trying to preserve Lorentz invariance as much as possible
we demand
$\tilde{u}(p)=\Lambda_p\tilde{u}(0)$ and $\tilde{v}(p)=\Lambda_p\tilde{v}(0)$, obtaining
\beq \label{eq:Wprot}
W(p)= \Lambda_p W(0) \Lambda_p^{-1} .\eeq
Since $W(0)$ commutes with $\gamma_0$, the condition in eq.\eq{W1} for preserving the total cross sections is satisfied if
$U,V$ are U(2) rotations:
\beq \label{eq:WdW} U^\dagger U = \One = V^\dagger V.\eeq
Later computations will be simplified by noticing that this new physics can be
equivalently rewritten using the constant matrices $U,V$ acting on the spin indices
\beq \label{eq:tiltedspinors2} \tilde{u}_s(p) \equiv u_{s'}(p) U_{s's} ,\qquad
\tilde{v}_s(p) \equiv v_{s'}(p) V_{s's}^* .\eeq
Then the general Lorentz transformation of eq.\eq{LorentzW} reduces to
\beq U \to U' = D U D^{-1},\qquad V \to V' = D V D^{-1}\eeq
showing that $U$ and $V$ transform as rotations.
This means that $U,V\neq \mathbb{1}$ break rotational invariance and that
Lorentz invariance is broken too, but only by the Wigner rotations $D$ associated to boosts.
Boosts are broken in the following mild way: the reference frame $S$ is special because
in $S$ the breaking of rotational invariance has the simple form of eq.\eq{Wprot}.\footnote{Eq.\eq{Wprot} assumes that in one reference frame $S$ 
the factor $W(p)$  for generic momentum $p$ is related to its value as rest $W(0)$ as $W(p)= \Lambda_p W(0) \Lambda_p^{-1}$ .
We here discuss the analogous relation in a generic reference frame $S'$, showing that extra Wigner rotations appear.
In a different reference frame $S'$,  related to $S$ by a Lorentz transformation $\Lambda$, the momentum $p$ becomes $p' =\Lambda p$
and $W'(p') = \Lambda W(p) \Lambda^{-1}$, eq.\eq{LorentzW}.
In $S'$ one can similarly try writing $W'(p') =  \Lambda'_{p'} W'(0) \Lambda_{p'}^{\prime -1}$.
If the theory were Lorentz invariant, $W'(0)$ would be a $p'$-independent matrix.
 The above equations show that $W' (0)= R W(0) R^{-1}$
where $R=\Lambda_{p'}^{-1} \Lambda \Lambda_p$ is a $p'$-dependent Wigner rotation.}



\medskip

Eq.\eq{CPT} implies that parity P, charge conjugation C, and time reversal T act on $U,V$ as
\beq
\begin{array}{llll}
\hbox{P:} & U \xrightarrow{\rm P}  U & V \xrightarrow{\rm P}  V  \cr
\hbox{C:}  &U \xrightarrow{\rm C} V & V \xrightarrow{\rm C} U  \cr
\hbox{T:}  &
U \xrightarrow{\rm T} \epsilon U^* \epsilon^{-1} & V \xrightarrow{\rm T} \epsilon V^* \epsilon^{-1} 
\end{array}\label{eq:CPT2}
\eeq
Notice that  $\epsilon U^* \epsilon^{-1} =U$ if $U$ is a $\SU(2)$ rotation, and similarly for $V$.
In the following, we will assume that $U$ and $V$ are SU(2) rotations, so that T is conserved and $W =\diag(U,V)$.
In order to break Lorentz invariance minimally we will consider two sub-cases, C-even and C-odd, 
analogous to the ones in section~\ref{SO21}:
\begin{enumerate}
\item $V=U$ are two {equal} rotations with angle $\delta/2$ around the same axis $\mb{n}$. 
Then $W(0)$ is a rotation in 4-dimensional spinor space, and conserves C, as well as P and T.
A rotation remains $W(0)=\diag(U,U)$ also in the Weyl basis.
It can be written as
\beq \label{eq:W0rot}
W(p) =\exp(i \delta\, [\slashed{n}_p,\slashed{p}]\gamma_5 /8M_p)
\eeq
which is a rotation with angle $\delta/2$ around the boosted 
$n^\mu_p = (\Lambda_p)^\mu_{~\nu} n^\nu$.
For $p$ at rest and $\mb{n}$ along the $z$ axis it reduces to 
$W(0)=\diag (e^{i\delta/4},e^{-i\delta/4},e^{i\delta/4},e^{-i\delta/4})$ as in eq.\eq{WSO21a},
having defined $M_p$ in the same way.


\item $V =U^{-1}$ are two {opposite} rotations with angle $\delta/2$ around the same axis $\mb{n}$.
The assumed $W =\diag(U,U^{-1})$  in the Dirac basis can be written in a basis-independent way as 
\beq  \label{eq:Wn}
W(p) =\exp(-i \delta\, \slashed{n}_p\gamma_5/4).
\eeq
For $p$ at rest and $n_p=n$ along the $z$ axis it reduces to 
$W(0)=\diag (e^{i\delta/4},e^{-i\delta/4},e^{-i\delta/4},e^{i\delta/4})$ as in eq.\eq{WSO21b}.
P and T are conserved, and the parameter $\delta$ is odd under C, CP and CPT  (see table~\ref{tab:CPT}).
\end{enumerate}
These $W(p)$ differ from the analogous one in eq.s\eq{WSO21a} and \eq{WSO21b}
because $n$ has been promoted into $n_p$.
As a result all boosts are broken, but only by the Wigner matrices that rotate $n^\mu$.
The Lorentz group is broken to SO(2), describing rotations around $\mb{n}$.


\subsection{Lorentz broken to its maximal sub-group SIM(2)}\label{SIM}
The maximal Lorentz sub-group $H={\rm SIM}(2)$ is spanned by the 4 generators~\cite{wiki,hep-ph/0601236}
\beq J_z,\qquad K_z, \qquad T_1= K_x+J_y,\qquad T_2=K_y-J_x. 
\eeq
The light-cone combinations $T_1$ and $T_2$ commute, forming a group of 2-dimensional translations.
The SIM(2) sub-algebra closes as 
\beq\label{eq:SIM2}
 [T_1, K_z]=i T_1,\qquad [T_2, K_z]=i T_2,\qquad
[T_1, J_z]=-i T_2, \qquad [T_2,J_z]=i T_1.\eeq
$H$ is denoted as SIM(2) because this Lie algebra is equivalent to the one of similitude transformations in 2 dimensions.
SIM(2) contains all boosts $K_{x,y,z}$, so a generic time-like 4-vector can be rotated to its rest frame
(at the price of an extra Wigner rotation performed by $K_{x,y}$ in $T_{1,2}$)
and the speed of light $c$ remains universal~\cite{hep-ph/0601236,1008.0436}.
The vector 
\beq n^\mu=(1,0,0,1) \eeq
is invariant under the sub-group spanned by $T_1,T_2,J_z$  and transforms multiplicatively under $K_z$.
Thereby Lorentz-breaking SIM(2)-invariant theories can be written from non-local
ratios of terms containing the same power of $n$~\cite{hep-ph/0601236}.
This can be immediately generalized to $n^\mu = (1,\mb{n})$.
Notice that $n^\mu$ differs from the previous sections~\ref{SO21} and~\ref{rot},
where $n^\mu= (0,0,0,1)$ resulted in different sub-groups.
We present three concrete SIM(2)-invariant theories, corresponding to 3 choices of $W(p)$.

\begin{enumerate}
\item[0.] Assuming that $W$ is a boost along $\mb{n}$ with parameter $\delta$ gives
\beq \label{eq:SIMW0}
W(p)=\exp \left(\frac{\delta}{4}\frac{[\slashed{n},\slashed{p}]}{n\cdot p} \right)
,\qquad
\tilde{\cal K} \equiv
\overline{W}(\slashed{p}-M)W=e^{\delta} \slashed{p}-M- \frac{p^2 \, \slashed{n}}{p\cdot n} \sinh\delta .
\eeq 
This analytic  $W(p)$ satisfies $\overline{W} W=1$, so $\tilde{{\cal K}}^2 = p^2 - M^2$ reproducing the standard relativistic dispersion relation.
The modified Dirac operator $\tilde{\cal K}$ generalises the theory proposed in~\cite{hep-ph/0605036} for $M=0$.
Since $\tilde{\cal K} \neq {\cal K}$ this theory affects spin-summed cross sections.

\item[1.] A different SIM(2)-invariant  theory is obtained assuming
that $W(p)$ is a rotation with angle $\delta/2$ around
$\mb{n}$ acting in the same way on particles and anti-particles
(analogously to the C-even cases 1.~in sections~\ref{SO21},~\ref{rot}):
\beq \label{eq:SIMW1}
W(p)=\exp \left(i\frac{\delta}{4}\frac{[\slashed{n},\slashed{p}]\gamma_5}{n\cdot p} \right),\qquad
W(0)=\diag (e^{i\delta/4},e^{-i\delta/4},e^{i\delta/4},e^{-i\delta/4}).\eeq
This analytic $W(p)$ satisfies eq.\eq{W1} leaving the Dirac operator $\tilde{\cal K}={\cal K}$, and spin-summed tree-level cross sections unchanged.

\item[2.] One more SIM(2)-invariant theory is obtained assuming 
that $W(p)$ is a rotation with angle $\delta/2$ around $\mb{n}$ acting in the opposite way on particles and anti-particles
(analogously to the C-odd cases 2.~in sections~\ref{SO21},~\ref{rot}):
\beq\label{eq:SIMW2}
W (p)=  \exp\left[- i \frac{\delta}{4}\left( \frac{\slashed{n}\, M_p}{n\cdot p} - \frac{\slashed{p}}{M_p}\right)\gamma_5 \right]
,\qquad
W(0)=\diag (e^{i\delta/4},e^{-i\delta/4},e^{-i\delta/4},e^{+i\delta/4}).\eeq
The second term in the exponent removes the `time-like' component of $n$ parallel to $p$.
We again defined $M_p$ as in eq.\eq{WSO21a} such that $W(p)=W(-p)$ has a common limit
at rest, $W(0)$.
This SIM(2) theory satisfies eq.\eq{W1}.
\end{enumerate}

\medskip

Eq.\eq{CPT} describes how the discrete symmetries C, P, and T  act on $W(p)$.
Table~\ref{tab:CPT} lists how the various $W(p)$ conserve or break the space-time discrete symmetries.
P, T, CP and CT must be broken in all SIM(2) theories, as
their conservation would enlarge SIM(2) to the full Lorentz group~\cite{hep-ph/0601236}.
CPT  is conserved in all analytic  SIM(2) theories~\cite{hep-ph/0601236}.
The theories 0.\ and 1.\ conserve C (the new physics couples in the same way to particles and anti-particles),
while theory 2.\ breaks C (particles and anti-particles couple in opposite ways to the new physics).

\section{Tree level effects}\label{tree}
We here discuss how the tree-level processes are affected by our new physics $W(p)$,
and we choose it to realise our phenomenological goal:
modifying entanglement in $pp \to t\bar t$ without affecting the spin-summed scatterings and decays.
Motivated by this phenomenological application
we assume that the $\Psi \to \tilde\Psi = W \Psi$ transformation is operated on the QCD top quark interactions,
while leaving untouched all other top electro-weak and Yukawa interactions and all other SM particles.
In this way $W$ becomes physical, describing a misalignment of top spin between QCD vs EW interactions.
As top quarks are much heavier than the QCD scale, this new physics effect in top spins is negligibly diluted by hadronization.
We can work in the basis where only the vertex of eq.\eq{QCDvertex} gets modified.

%
%
%


\smallskip

\begin{figure}[t]
$$\includegraphics[width=0.6\textwidth]{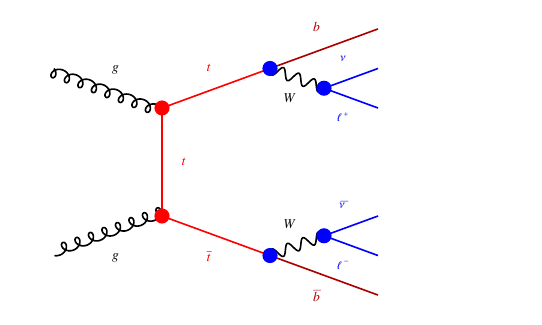}$$
\caption{\em\label{fig:FeynEntanglementLHC} A Feynman diagram
for top entanglement at the LHC, with QCD (red) and EW (blue) vertices modified by different $W(p)$ factors.}
\end{figure}

We compute the dominant partonic process, $t \bar{t}$ productions from gluons $g$.
The tree-level amplitude for 
$g^a_{\lambda_1}(p_1)g^b_{\lambda_2}(p_2)\to t_{i s_1}(q_1)\bar{t}_{j {s_2}}(q_2)$
($a,b$ are adjoint indices, $i,j$ are color indices, $\lambda_{1,2}$ are gluon helicities,
$s_{1,2}$ are top spins), obtained summing the usual $s,t,u$-channel diagrams, is
\begin{eqnarray} \mathscr{A}&=& -ig_3^2 \, \epsilon^{\lambda_1}_\mu(p_1) \,\epsilon^{\lambda_2}_\nu(p_2)\,
\overline{\tilde{u}}_{s_1}(q_1) \bigg\{
 \gamma^\mu  \frac{ (T^a T^b)_{ij}}{\slashed{q}_1-\slashed{p}_1 - M}  \gamma^\nu  +
  \gamma^\nu \frac{ (T^b T^a)_{ij}}{\slashed{q}_1-\slashed{p}_2 - M}  \gamma^\mu  +
  \label{eq:Aggtt}
 \\
&&+ i \frac{f^{abc}}{s} T^c_{ij}
[\eta^{\mu\nu}(\slashed{p}_1-\slashed{p}_2)^\rho + \gamma^{\nu}(p_1+2p_2)^\mu - \gamma^{\mu}(2p_1+p_2)^\nu] 
\bigg\}  \tilde{v}_{s_2}(q_2). \nonumber
 \end{eqnarray}
This is the standard SM amplitude with spinors replaced by tilted spinors
\beq \label{eq:tiltedtop}
\overline{\tilde{u}}_{s_1}(q_1) = \overline{u}_{s_1}(q_1) \overline{W}(q_1) ,\qquad
\tilde{v}_{s_2}(q_2) = W(-q_2)  v_{s_2}(q_2).\eeq
Squaring the amplitude and
summing over spins 
gives factors that reproduce the standard spin sums
\beq 
\sum_s {\tilde{u}}_{s}(q)   \overline{\tilde{u}}_{s}(q)
= W(q) ( \slashed{q}+ M)\overline{W}(q) =
  \slashed{q} + M =  \sum_s {{u}}_{s}(q)   \overline{{u}}_{s}(q)
\eeq
\beq 
\sum_s {\tilde{v}}_{s}(q)   \overline{\tilde{v}}_{s}(q)
= -W(-q) (- \slashed{q}+ M)\overline{W}(-q) =
  \slashed{q} - M =  \sum_s {{v}}_{s}(q)   \overline{{v}}_{s}(q)
\eeq
since we assumed that $W$ satisfies eq.\eq{W1}.
So the cross section summed over spins is not affected by $W$.
The same happens for the other $q\bar{q}\to t\bar{t}$ partonic processes.
So the spin-summed $pp\to t\bar t$ cross section at the tree level remains the same as in the SM,
while the top spin structure is affected.
This is manifest in observable processes such as fig.\fig{FeynEntanglementLHC}
where the spin-summed differential rates for  $gg \to t\bar{t}$ and for $t \to b \bar\ell\nu$ remain separately unmodified,
while the $gg \to b\bar \ell\nu \bar b \ell\bar\nu$ differential rates are modified.

\subsection{Spin correlation and entanglement in $pp \to t \bar {t}$}
To discuss how the spin structure is modified, we now introduce the standard formalism.
Multiple partonic processes contribute to $pp \to t(q_1,s_1) \bar t (q_2,s_2)$ processes,
each one with amplitudes $\mathscr{A}(q_{1,2},s_{1,2})$ that depend on the $t$ and $\bar{t}$ momenta $q_i$ and spins $s_i$.
As usual, the squared amplitudes are summed over processes, integrated over their parton densities, 
averaged over initial-state spins.
However, we do not sum over final-state spins, as we want to retain this information.
The scattering rates get thereby described by a $4\times 4$ matrix in spin space.
It is convenient to write this matrix as the usual total 
differential cross section $d\sigma(q_1, q_2)/dt$ (summed over final-state spins)
times a $4\times 4$ density  matrix $\rho(q_1,q_2)$ in $s_1,s_2$ space
normalised as $\Tr\,\rho=1$
that describes the spin composition of each final $t(q_1)\bar t(q_2)$ state.
Furthermore, $\rho$ is usually parameterised in terms of $2\times 2$ Pauli matrices $\sigma_i$ 
as 
\beq \label{eq:rho}
\rho =\frac{1}{4}(\One\otimes \One + (\vec\sigma\cdot \vec{S}_1)\otimes \One+
\One\otimes (\vec\sigma\cdot\vec{S}_2) + C_{ij}\, \sigma_i \otimes\sigma_j).\eeq
In this way the first $\One\otimes \One$ term accounts for the usual total cross section;
$\vec{S}_{1,2}$ tell the average spins of $t$ and of $\bar{t}$;
the spin-correlation matrix $C$  can contain entanglement.
For example the spin singlet $t\bar t$ state corresponds to $C = -\diag(1,1,1)$.
The spin triplet state along the $z$ axis corresponds to $C=\diag(1,1,-1)$.

The new physics introduced in section~\ref{th} 
modifies the SM amplitudes as in eq.s\eq{Aggtt}, \eq{tiltedtop}.
This can be conveniently rewritten by defining two $2\times 2$ matrices $U(q_1)$ and $V(q_2)$ acting on spin indices $s_{1,2}$
as:
\beq \tilde u_{s_1}(q_1) = \Lambda_{q_1} (\Lambda_{q_1}^{-1} W(q_1) \Lambda_{q_1}) u_{s_1}(0)\equiv
\Lambda_{q_1} U_{s_1s_1'}(q_1) u_{s'_1}(0) = U_{s_1s'_1}(q_1) u_{s'_1}(q_1)  \eeq
and similarly for $v$.
In the theory of section~\ref{rot} the matrices $U$ and $V$ do not depend on momenta in frame $S$.
In the theories of sections~\ref{SO21} and \ref{SIM}, $U(q_1)$ and $V(q_2)$ are rotations along a rotated $n$, related
to $U(0)$ and $V(0)$ by Wigner rotations.
In the density matrix language, this means that $\rho$ of eq.\eq{rho} gets modified into a $\tilde\rho$
where $\mb{S}_{1,2}$ and $C$ keep their SM values, and
the Pauli matrices for $t$ and for $\bar{t}$ get tilted as 
\beq \sigma_i \to \tilde{\sigma}_i = U^\dagger \sigma_i U \equiv R_{Uij}\sigma_j,\qquad
\sigma_i \to \tilde{\sigma}_i = V^\dagger \sigma_i V \equiv R_{Vij}\sigma_j.\eeq
The above equation means that
tilted Pauli matrices can be converted to the usual Pauli basis by noticing
that the $U$ and $V$ rotations in spin space induce
 rotations in space, described by the $3\times 3$ matrices $R_U$ and $R_V$. 
 So the $\tilde{\rho}$ density matrix can be rewritten in the standard Pauli basis as
 \beq \tilde{\rho}=
\frac{1}{4}(\One\otimes \One + 
({\vec{\sigma}}\cdot \tilde{\vec{S}}_1)\otimes \One+
\One\otimes ({\vec{\sigma}}\cdot\tilde{\vec{S}}_2) + \tilde{C}_{ij}\, \sigma_i \otimes \sigma_j)
\eeq
where the $t$ and $\bar t$ spins and their correlation matrix get rotated as 
\beq \label{eq:Ctilde}
\tilde{\mb{S}}_1 = R_U^T \cdot  \mb{S}_1,\qquad
\tilde{\mb{S}}_2 = R_V^T \cdot  \mb{S}_{2}   ,\qquad
\tilde{C} = R_U^T \cdot C \cdot R_V.\eeq 
As expected, the new physics amounts to just rotations of the arbitrary quantization axis, separately for $t$ and $\bar t$.
These rotations become physical when we assume that  the QCD interactions of top quarks
are modified by the $U$ and $V$ matrices, while the
top weak interactions (responsible for $t$ decays) are not modified.
In this way $U$ and $V$ modify, in particular, the angular distributions of leptons $\ell$ in
processes where tops are produced via QCD interactions and decay via weak interactions, $t \to b \bar\ell\nu$.

\subsection{Comparison with Large Hadron Collider data}

The lepton angular distributions in the $t\bar{t}$ rest frame have been used by LHC experiments to measure
the trace 
\beq D \equiv \frac{\Tr\,C}{3} =  \frac{2}{3}\langle {S}^2 \rangle  - 1\eeq 
of the $pp \to t\bar t$ spin correlation matrix, sensitive to the total spin $\vec{S}=\vec{S}_1+\vec{S}_2$ of the $t\bar t$ pair. 
In order to modify the $D$ observable from its SM value
we need to assume that our new physics introduces a relative rotation $U\neq V$ between $t$ and $\bar t$.
The $\tilde{C}$ matrix is no longer symmetric, implying that CP is broken~\cite{hep-ph/9312210,1212.4888,1508.05271}.

\medskip

The $D$ observable is sensitive, in particular, to spin entanglement.
Entanglement is necessarily present if $D < -1/3$~\cite{Afik:2020onf} i.e.\ if $\med{{S}^2} 
<1$.
ATLAS confirms top entanglement by measuring, with $140/{\rm fb}$ of integrated luminosity at $\sqrt{s}=13\TeV$~\cite{ATLAS},
\beq \label{eq:ATLASD}
D = -0.547\pm 0.021_{\rm syst}  \pm 0.002_{\rm stat}\qquad \hbox{for}\qquad
340\GeV< m_{t\bar t} < 380\GeV.
\eeq
The ATLAS analysis restricts the $t\bar{t}$ invariant mass $m_{t\bar t}$ 
to non-relativistic values because,
in this limit, the SM predicts that the dominant
$gg\to t\bar t$ channel  
gives a maximally entangled spin-singlet state, which corresponds to $D=-1$.
Taking into account the finite bin size and the subdominant $q\bar q \to t\bar t$ channel,
the SM prediction~\cite{1212.4888,hep-ph/9312210, Uwer:2004vp}
computed using the {\sc Powheg} + {\sc Pythia} modelling 
is $D_{\rm SM}= -0.470\pm 0.018_{\rm syst}  \pm 0.002_{\rm stat}$~\cite{ATLAS}.
A second computation, based on the {\sc Powheg} + {\sc Herwig7} modelling, 
 yields an even higher central value~\cite{ATLAS}. 
Performing a simple parton-level approximation we find\footnote{This is computed as
$C_{\rm SM} = \med{\omega^T C_{\rm hel} \omega}$ where $\med{\cdots}$ denotes the average
over the phase space of final-state tops with the experimental cuts, 
$C_{\rm hel}$ is the correlation matrix computed in the helicity basis,
and $\omega = R_{13}(\theta) R_{12}(\pi) R_{12}(\phi)$ is the rotation matrix
that connects the helicity basis to fixed $x,y,z$ axis,
and $\theta,\phi$ are the usual scattering angles in the center of mass frame.
}
\beq \label{eq:CSM}
C_{\rm SM} \approx-\diag(0.54,0.54,0.18) \qquad \hbox{i.e.}\qquad D_{\rm SM}\approx -0.42.
\eeq

The measured $D$ is about $2.8$ standard deviations below the SM predictions.
We ignore this minor anomaly, possibly due to neglected non-relativistic QCD processes such as $t\bar{t}$
bound state effects which affect the production of non-relativistic $t\bar t$ events~\cite{Kiyo:2008bv,2401.08751}.
Different computations agree with each other and with data at higher $m_{t\bar{t}}$ away from the threshold,
where data does not establish top entanglement~\cite{ATLAS}. 
The CMS experiment too established entanglement~\cite{2406.03976}, after having selected non-relativistic $t\bar t$ pairs. 
The value obtained by CMS with 
36.3/fb of integrated luminosity at $\sqrt{s}=13$~TeV is~\cite{2406.03976}
\beq \label{eq:CMSD}
D = -0.480^{+0.020}_{-0.023}{\,\rm (syst)} ^{+0.016}_{-0.017}{\,\rm (stat)}\qquad \hbox{for}\qquad
345\GeV< m_{t\bar t} < 400\GeV.
\eeq
It is in good agreement with the SM prediction, 
which includes an estimate of the hypothetical toponium contribution.
The SM prediction is computed in three different ways, 
using i)~{\sc Powheg} + {\sc Pythia8},
ii)~{\sc MG5\_aMC@NLO} + {\sc Pythia8}, 
and iii)~{\sc Powheg} + 
{\sc Herwig$++$}.
From our simple parton-level approximation, assuming $345\GeV< m_{t\bar t} < 400\GeV$ and $\sqrt{s} = 13$~TeV, we find
\beq \label{eq:CSM2}
C_{\rm SM} \approx-\diag(0.49,0.49,0.13) \qquad \hbox{i.e.}\qquad D_{\rm SM}\approx -0.37.
\eeq

Our new physics effects that modify entanglement depend on the momenta of the produced tops.
The operators $W(p)$ of section~\ref{SIM} predict an enhancement for relativistic tops aligned to $\mb{n}$.
For simplicity we focus on the ATLAS sample of non-relativistic tops, eq.\eq{ATLASD}.
Then, in order to significantly modify the $D$ observable, we focus 
on the theories discussed at points 2.~of sections~\ref{SO21}, \ref{rot} and \ref{SIM},
such that $U$ and $V$ are approximatively {\em opposite} rotations  with angle $\delta/2$ around the same unit vector $\mb{n}$.
Inserting $U = V^{-1}$  and thereby
$R_V = R_U^T \equiv R$ in  eq.\eq{Ctilde}, it simplifies into
\beq \tilde{C} = R\cdot C_{\rm SM}\cdot R.\eeq
A fit to ATLAS (CMS) data implies the bound
$|\delta|\lesssim 0.6$ (0.7)
at $3\sigma$ 
with $\mb{n} = (0,0,1)$. 
For perpendicular directions, the bound is $18\%$ weaker, therefore the proper averaging over various rotating LHC orientations is only a minor correction.
Without loss of generality, we can consider a rotation along the spin-quantization $z$ axis.
It corresponds to a diagonal matrix of phases in the Dirac basis
as in eq.s\eq{WSO21b},\eq{Wn} and\eq{SIMW2},
$W(0)=\diag (e^{i \delta/4}, e^{-i \delta/4}, e^{-i \delta/4}, e^{i \delta/4})$.
In its presence,
$gg$ collisions produce, in the non-relativistic limit, $t\bar{t}$ pairs with entanglement modified into 
\beq\label{eq:delta} \kk{\delta}= \frac{e^{-i\delta/2} \kk{t_\uparrow \bar{t}_\downarrow}- e^{i\delta/2} \kk{t_\downarrow \bar{t}_\uparrow}}{\sqrt{2}}.\eeq
The density matrix
$\rho=\kb{\delta}{\delta}$ corresponds to the spin correlation matrix
\beq \tilde{C} =-\begin{pmatrix}
\cos\delta &- \sin\delta & 0 \cr \sin\delta & \cos\delta & 0 \cr 0 & 0 & 1
\end{pmatrix}.
\eeq
The value of $\tilde{D}=\frac13 \Tr\tilde{C}=-\frac13 - \frac23\cos\delta$ is increased by $\delta\neq 0$.
This shows that $\delta$ affects the entanglement measured by ATLAS/CMS.
In the limit of a large random $\delta$, quantum coherence is lost giving classical probabilities with no entanglement.


\section{Loop effects}\label{loop}
In general, new physics in entanglement is expected to affect quantum loops.

However, it is not possible to compute in a model-independent way how loops are affected by a generic loss of entanglement.
The reason is that  a complete loss of entanglement turns amplitudes into classical probabilities,
but loops are amplitudes (such as the loop correction to a particle mass).
Specific theories of new physics in entanglement, such as the one we proposed,
are needed to study how quantum loops are affected.

To make the discussion concrete, we focus on our phenomenological application.
Presumably, quantum loops spread the new physics we introduced into top quarks
among all sectors of the theory, while respecting
the residual symmetries (the Lorentz sub-groups SO(2), SO(2,1), SIM(2) are considered in section~\ref{th})
and possibly the form of $W(p)$.
A similar outcome was found in a different context: introducing the simplest form of Lorentz-breaking effects 
--- different speeds-of-light $c_p$ for different particles $p$ ---
the $c_p$ obey a closed system of Renormalisation Group Equations, computed at one loop in~\cite{1003.2364}.

We assumed that top quarks are affected
by a non-local $W(p)$ rotation in spin space that 
affects differently the top QCD and EW couplings.
So deviations from the Standard Model can arise in processes mediated by top loops
that involve both QCD and EW interactions.
Corrections often arise at two or more loops, 
and our postulated $W(p)$ breaks rotational invariance.
These two characteristics render a technical challenge in computing loop corrections
to EW physics, Higgs physics, flavour physics, and low-energy physics.

As we will only estimate the loop effects, we only briefly discuss their gauge (in)variance.
The modified vertex of eq.\eq{QCDvertex}, when rewritten in coordinate space, corresponds to a non-local interaction
$g_3 \overline{\tilde{\Psi}} (x_1) \gamma^\mu T^a \tilde{\Psi}(x_2) g^a_\mu(x_3)$ at $x_1\neq x_2\neq x_3$.
The non-locality breaks weak local gauge invariance, as a gauge transformation rotates top quarks with generic different phases at $x_1$ and $x_2$.
One exception is the $p$-independent $W$ of eq.\eq{WSO21b} that enforces $x_1=x_2$. Other exceptions exist.
As long as non-locality only happens on small scales, the consequent breaking of gauge invariance could too be phenomenologically harmless.
Otherwise, non-locality and local gauge invariance are, in general, compatible.\footnote{For example, ordinary
loop-level effective actions are non-local and gauge-invariant.
The effective theory we are considering could arise, at fundamental level, by assuming that top quarks couple to Lorentz-breaking
particle(s) with electroweak and strong gauge charges different from the top itself.}
The Noether procedure should allow to modify our theory making it
gauge invariant by adding suitable interactions with the SM gauge vectors order by order in perturbation theory
in the gauge couplings.
Explicit non-local gauge-invariant actions are known in the special case of lattice non-locality.
More general examples have been formulated in~\cite{Partovi:1982yx}:
it involves generalising $W(i \partial)$ into ${\rm P}\, W(i D)$, where $D$ is the gauge-covariant derivative
and the path ordering P deals with their non-commutation.
We do not explore more this possibility.


\begin{figure}[t]
$$\raisebox{-0.03\textwidth}{\includegraphics[width=0.3\textwidth]{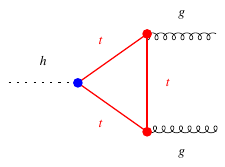}}\quad
\includegraphics[width=0.3\textwidth]{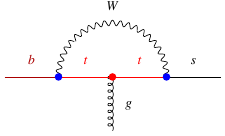}\quad
\includegraphics[width=0.3\textwidth]{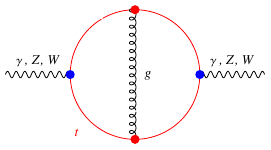}$$
\caption{\em\label{fig:FeynEntanglementLoop} Sample of Feynman diagrams of observed top loops
modified by our new physics in entanglement as both QCD (red) and EW (blue) vertices appear.}
\end{figure}

\subsection{Higgs physics}
Concerning Higgs physics, the SM predicts that $\Gamma(h \to gg)$, $\Gamma(h\to \gamma\gamma)$, $\Gamma(h\to Z\gamma)$
arise at one loop, mediated by top quark loops (as well as by EW loops).
LHC experiments find that these rates agree with the SM within a few tens of $\%$ precision.
Our new physics first affects $h\to \gamma\gamma$ and $h\to Z\gamma$ at two loops, 
so we estimate that experimental bounds are satisfied for $\delta \sim 1$.
$\Gamma(h \leftrightarrow gg)$ is less precisely measured from the $pp\to h$ production rate,
and it's affected at one loop.
The expression for the diagram in fig.\fig{FeynEntanglementLoop} left
can be derived from the tree-level expression for $gg \to t(q_1)\bar t(q_2)$, eq.\eq{Aggtt},
closing the top loop with an extra top Yukawa interaction, given in the SM by $y_t \One$ where $\One$ is the unit matrix in spinor space.
The overall effect of our new physics 
affecting differently QCD vs EW top couplings
is replacing $y_t \One$ by
\beq  y_t\, W(q_1) \overline{W}(-q_2) = y_t \, W(k) \overline{W}(k-p),\eeq 
with $k$ being the loop momentum 
and $p$ the momentum of the Higgs.
The new physics correction trivially vanishes in the limits $p=0$ and $\delta=0$. 
Expanding the $W(q)$ matrices in $\delta \ll 1$ and calculating the Dirac trace, we find that terms linear in $\delta$ do not contribute to the decay width due to $\gamma_5$ accompanying each $\delta$ factor.
This allows to estimate the leading relative correction to $h \leftrightarrow gg$ as
$\Delta \Gamma/\Gamma \approx \lambda\delta^2 M^2_h/M^2_t $, where the factor $\lambda$ arises from the loop integral. We find $\lambda=3/4$ for $W(q)$ in the form (\ref{eq:SIMW1}) and 
$\lambda \approx 1.3$ for the operator (\ref{eq:SIMW2}).
ATLAS and CMS measure the Higgs production cross-section in the gluon fusion channel with~$21\%$ precision\cite{2207.00092,2207.00043} at $3\sigma$ level, that for $\lambda\approx 1$ results in the constraint $|\delta|<0.6$ comparable to the one obtained from the entanglement observable.
We note that there is no correction to this process for the $p$-independent $W$ of eq.\eq{WSO21b}.

\subsection{Electroweak precision physics}
Coming to EW precision physics,
the SM predicts that two EW precision parameters have, at one loop, a strong quadratic sensitivity to the top quark mass:
the $ \rho\equiv {M_W^2}/{M_Z^2 \cos^2\theta_{\rm W}}$ parameter and the 
$\epsilon_b$ parameter that describes the $Z b_L \bar b_L$ coupling.
These parameters are first affected at two loops (e.g.~fig.\fig{FeynEntanglementLoop} right),
as one-loop diagrams involve either QCD or EW interactions, but not both.
So the new physics effects are estimated to be below current bounds.
In the same way, new physics corrections to the top quark propagator first arise at two loop level.
If instead new physics affects bottom quarks too, the $\epsilon_b$ parameter is affected at one loop.

\subsection{Flavour physics}
Coming to flavour physics, measurements of rare decays and oscillations of mostly $K$ and $B$ mesons 
allowed to test the coefficients of various effective operators that, in the SM, receive contributions from 
top quarks at one loop level, via box and penguin diagrams. 
The new physics we introduced mostly affects such coefficients at one higher order than leading SM effects,
with the exception of QCD penguins.
These first arise and are affected at one loop (fig.\fig{FeynEntanglementLoop} middle).
Since an external gluon is involved, such operators suffer larger QCD uncertainties.
Flavour physics currently is $2-3$ times less sensitive to top quark loops than EW precision tests
(see for example the determination of the top quark mass~\cite{1508.05332}).
An exception would arise if the rotation-breaking and/or CP-breaking new physics gives rise to qualitatively new effects,
such as chiral enhancements.
Qualitatively new effects are however better tested moving from particle-physics experiments to
low-energy probes, which we discuss next.


\subsection{Low-energy physics}
The new physics we introduced  affects the photon propagator $\Pi_{\mu\nu}(k)$ first at 2 loops
(a sample diagram is shown in the right panel of fig.\fig{FeynEntanglementLoop})
and the electron propagator at 3 loops.
As we are not able to compute these multi-loop diagrams, we limit our discussions to symmetry arguments.

\medskip

If Lorentz is broken to its smaller
SO(2) or SO(2,1) subgroups (as in sections~\ref{SO21},~\ref{rot}), 
these residual symmetries have little protective power.
In the worst case, loops induce effective operators of the generic form $k_{\mu\nu\alpha\beta} F^{\mu\nu}F^{\alpha\beta}$,
with dimension-less coefficients arising at order $k_{\mu\nu\alpha\beta}  \sim e^2 g_3^2/(4\pi)^4$.
Some combinations of coefficients $k$ 
are experimentally constrained down to $10^{-34}$ level by 
observations of the polarization of sources at cosmological distance~\cite{0801.0287,2104.00238}.
The combination $\delta c$ of $k_{\mu\nu\alpha\beta} $ coefficients that controls the speed of light 
(compared to the speed of ultra-relativistic electrons or protons)
is constrained as $\delta c \lesssim 10^{-15}$~\cite{0801.0287}.
The SO(2,1)-invariant theory of eq.\eq{WSO21b} can be computed finding that it is problematic,
as the $p$-independent $W$ modifies the photon vertex $\gamma_\mu$ rotating $\gamma_3$ into $\gamma_5$.
The SO(2)-invariant theories of section~\ref{rot} are not easily computed; 
one might hope that loops preserve their assumed structure  
(one special frame where new physics induces fixed spin rotations),
giving a rotation $R$ of photons and their spin,
$\delta \Pi_{\mu\nu}=R_{\mu\mu'}R_{\nu\nu'}\Pi_{\mu'\nu'}^{\rm SM}(Rk)$.
This would be redefined away from the photon propagator 
(thereby bypassing bounds on it), leaving effects only in photon interactions.

\medskip If Lorentz is broken to its maximal sub-group SIM(2) (as in section~\ref{SIM}), 
this residual symmetry forbids most unwanted effects and only allows for a photon mass-like term
(of the form $m_\gamma^2 (n_\mu  F^{\mu\nu}/n^\alpha \partial_\alpha)^2$,
constrained to be $m_\gamma \lesssim 10^{-18}\eV$~\cite{0904.2065,1305.1577,1306.1941})
and for a spin-dependent electron mass (of the form 
$m_e \Delta m_e \bar\Psi \slashed{n}/(n\cdot\partial) \Psi$,
constrained to be $\Delta m_e \lesssim 10^{-20}\eV$~\cite{hep-ph/0611049}).
In view of the derivatives at the denominator, such terms should arise from 
loop integrals that develop IR divergences in the limit of small external momenta.
The SIM(2)-invariant theory of eq.\eq{SIMW0} gives an IR-divergent modified Dirac propagator, 
so that the photon mass could arise at one loop. 
An explicit computation found that the photon remains massless after regularising the IR divergences~\cite{hep-ph/0610202}.
We are instead interested in the SIM(2)-invariant theories that preserve spin-summed cross sections, eq.\eq{SIMW1} and\eq{SIMW2}.
The assumed exponential structure of $W(p)$ implies
that these theories don't introduce multiplicative IR divergences.
In particular, the Dirac propagator keeps its standard form, so there are no one-loop effects.
Two-loop effects only contain IR-enhanced spin rotations.
Dedicated computations are needed to see if a photon mass is generated at higher loops:
the non-local theory is likely non-renormalizable, in the sense that higher loop orders
likely modify our assumed benign forms of $W(p)$.

\medskip

In the worst case, low-energy effects could get naively suppressed by inverse powers of $\Lambda$,
by invoking the extra assumption that the new rotation-breaking physics 
is suppressed by some new physics scale $\Lambda \sim M_t$.
Such extra (unjustified?) assumption that Lorentz violation grows with energy is often adopted
in phenomenological studies, where $\Lambda $ is assumed to be the Planck mass.



\medskip

Finally, the strongest bound on Lorentz-violating effects (often ignored in the literature) comes from the vacuum energy.
The top-quark loop contribution to the energy-momentum tensor $T_{\mu\nu}$ is affected at 3 loops by our new physics.
Lorentz-breaking effects would complicate the apparent tuning of the vacuum energies down to its small observed value.

\section{Conclusions}\label{concl}
We presented  a theory of new physics that modifies spin entanglement without altering the total production or decay rates at the tree level.
As a phenomenological application, this new physics is relevant for tests of $t\bar t$ entanglement performed at the LHC,
as it affects the $t\bar t$ spin correlations and their entanglement,
without affecting the spin-summed differential QCD cross sections for $pp \to t\bar t$  top quark production,
nor the tree-level EW top differential decay rate summed over spins.

In order to accomplish this, we introduced an unconventional theoretical  ingredient: 
a rotation-breaking misalignment $\Psi(x) \to W(i\partial )\Psi(x)$ between the
QCD and EW interactions of top quarks. 
As a result, the polarizations of top quarks with momentum $p$ undergo a $W(p)$ rotation in spinor space
when both QCD and EW interactions are involved.
A significant new physics effect in the specific entanglement observable tested by ATLAS~\cite{ATLAS} and CMS~\cite{2406.03976}
arises if the breaking of rotational invariance couples in opposite ways to top and anti-top quarks.

\medskip

In general, one expects that new physics in top quark entanglement also affects processes affected by virtual
top quark loops. 
Such effects have been detected in Higgs physics, 
EW precision observables, flavor physics, and low-energy experiments. 
In section~\ref{loop} we showed that, in various cases, our new physics first enters at higher loop order.
The reason is that both QCD and EW interactions must be involved in the loop to have a new physics effect.

The new physics we introduced implies a non-local violation of Lorentz invariance.
We attempted to keep this violation below experimental constraints by postulating that, at tree level, it exclusively influences top quarks. 
However, at loop level, this violation extends to other particles better tested than tops.
In order to alleviate these Lorentz-violating effects, we assumed specific benign forms for the $W(p)$ function,
that break Lorentz and the space-time discrete symmetries to different sub-groups, as listed in table~\ref{tab:CPT}.
\begin{itemize}
\item In section~\ref{rot} we break Lorentz almost completely but
adjust the tree-level $W(p)$ to minimise the physical impact of broken boosts.
One rest frame is special just because, in it, the new physics that rotates spin is not affected by Wigner rotations.

\item In section~\ref{SIM} we break Lorentz to
its maximal SIM(2) sub-group that contains all boosts (mixed with rotations):
this unbroken Lorentz sub-group is powerful enough that no rest frame becomes special.
The speed of light $c$ remains universal, and most unobserved Lorentz-violating effects are similarly controlled by the residual symmetry.
Our findings provide new realizations of (necessarily non-local) theories invariant under SIM(2).
\end{itemize}
However, the resulting theories are likely non-renormalizable: at some level
quantum corrections might transform our assumed $W(p)$ functions into a more problematic form.
In such a case, the applicability of the theories under consideration would become limited 
primarily to tree-level phenomenological applications, such as in the domain of top physics.



\paragraph{Acknowledgements.} 
We thank Daniele Barducci, Gino Isidori and Luca Marzola.

\end{document}